\pdfoutput=1
\documentclass[preprint,journal]{vgtc}

\usepackage{booktabs}
\usepackage{amsfonts}
\usepackage{amssymb}
\usepackage{amsmath}
\usepackage{algorithm,algpseudocode}
\usepackage{color}
\usepackage[title]{appendix}

\onlineid{0}
\vgtccategory{Research}
\ieeedoi{10.1109/TVCG.2026.3653317}
\preprinttext{\parbox{0.9\textwidth}{$\copyright$~2026 IEEE. Author's accepted version of the article published in IEEE Transactions on Visualization and Computer Graphics, available at: \href{https://doi.org/10.1109/TVCG.2026.3653317}{\color{blue}https://doi.org/10.1109/TVCG.2026.3653317}}}
\manuscriptnote{Citation: D.~Rohmer, K.~Salem, N.~Kalyanasundaram, and V.~Zordan, ``Dynamic Skinning: Kinematics-Driven Cartoon Effects for Articulated Characters,'' \emph{IEEE Transactions on Visualization and Computer Graphics}, vol.~32, no.~3, pp.~2684--2694, March 2026, doi: 10.1109/TVCG.2026.3653317.\\[0.5ex]
$\copyright$~2026 IEEE. Personal use of this material is permitted. Permission from IEEE must be obtained for all other uses, in any current or future media, including reprinting/republishing this material for advertising or promotional purposes, creating new collective works, for resale or redistribution to servers or lists, or reuse of any copyrighted component of this work in other works.}
\author{%
  Damien Rohmer$^{1}$,
  Karim Salem$^{2}$, Niranjan Kalyanasundaram$^{2}$, and
  Victor Zordan$^{2}$
}
\affiliation{\normalsize\sffamily
  $^{1}$LIX, Ecole Polytechnique, CNRS, Institut Polytechnique de Paris, France\\
  $^{2}$Clemson University, USA
}
\authorfooter{
  \item
  	E-mail: damien.rohmer@polytechnique.edu, ksalem@clemson.edu, niranjan@nkportfolio.com, vbz@clemson.edu.
  \item
  	Karim Salem and Niranjan Kalyanasundaram were with Clemson University at the time of this work.
}

\abstract{%
 We present an extension to traditional rig skinning, like Linear Blend Skinning (LBS), to produce secondary motions that exhibit the appearance of physical phenomena without need for simulation.  At the core of the technique, we call \emph{dynamic skinning}, is a set of deformers which offset position of individual vertices as a function of position derivatives and time.  Examples of such deformers create effects such as oscillation in response to movement and the appearance of wave propagation, among others. Because the technique computes offsets directly and does not solve physics equations, it is extremely fast to compute. It also boasts a high degree of customizability which supports a desirable artist workflow and fine level of control. Finally, we showcase the technique in a number of scenarios and make comparisons with the state of the art.
}


\teaser{
  \centering
  \includegraphics[width=1.0\linewidth]{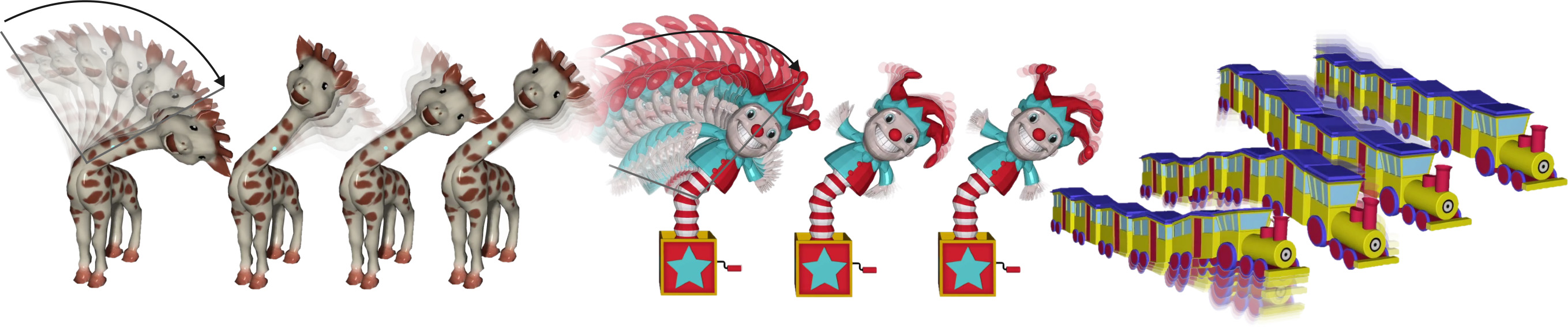}
  \caption{%
  	Dynamic skinning generates real-time oscillating behavior triggered by the skeletal motion, while automatically adapting to the character rig. The method is fully customizable to model global oscillation along the shape (see giraffe example), localized effects in some extremities (see clown example), or delayed deformation propagating along the skeleton (see train example).%
  }
  \label{fig:teaser}
}

\definecolor{turquoise}{cmyk}{0.65,0,0.1,0.1}
\definecolor{purple}{rgb}{0.65,0,0.65}
\definecolor{darkgreen}{rgb}{0.0, 0.5, 0.25}
\definecolor{darkred}{rgb}{0.8, 0.1, 0.0}
\definecolor{darkblue}{rgb}{0.0, 0.2, 0.5}
\definecolor{darkpink}{rgb}{0.67, 0.2, 0.416}
\definecolor{orange}{rgb}{0.9,0.5,0.0}
\definecolor{green}{rgb}{0.0,0.6,0.3}
\definecolor{black}{rgb}{0.0,0.0,0.0}
\definecolor{brown}{rgb}{0.9,0.7,0.1}

\definecolor{grey}{rgb}{0.5,0.5,0.5}
\definecolor{ocean}{rgb}{0.3,0.55,0.75}
\definecolor{lightgrey}{rgb}{0.6,0.6,0.6}	

\newcommand{\dron}{\color{purple}}
\newcommand{\droff}{\color{black}}

\newcommand{\vbz}[1]{\textcolor{black}{#1}}

\newcommand{\J}{\mathcal{J}}
\newcommand{\I}{\mathcal{I}}
\newcommand{\txt}[1]{\mbox{\tiny{#1}}}
\newcommand{\LBS}{\txt{LBS}}
\newcommand{\dLBS}{\txt{dLBS}}
\newcommand{\VS}{\txt{VS}}
\newcommand{\DS}{\txt{DS}}

\newcommand{\local}{\txt{local}}

\newcommand{\osc}{\,\txt{osc}}
\newcommand{\delay}{\,\txt{delay}}
\newcommand{\bend}{\txt{bend}}

\begin{document}

\title{Dynamic Skinning: \\ Kinematics-Driven Cartoon Effects for Articulated Characters}
\hypersetup{pdftitle={Dynamic Skinning: Kinematics-Driven Cartoon Effects for Articulated Characters}, pdfauthor={Damien Rohmer, Karim Salem, Niranjan Kalyanasundaram, Victor Zordan}, pdfsubject={IEEE Transactions on Visualization and Computer Graphics, 2026, DOI 10.1109/TVCG.2026.3653317 - Author's accepted version}, pdfkeywords={Animation, Skinning, Cartoon Effects, Real-Time}}

\firstsection{Introduction}

\maketitle

Producing expressive 3D cartoon-like character animation plays a major role in many animation productions and video games.
However, the process of generating and fine tuning these animations remains a significant time-consuming task for production studios. An essential element in crafting cartoon-like character animations lies in the ability to capture the character's liveliness through dynamic effects including exaggeration and others, grounded in the principles of animation used by animators for decades~\cite{illusionOfLife}.
Traditional production studios predominantly rely on manual techniques to create such expressive character animations. They employ carefully crafted rigs that are structured around skeletal articulation and so-called \emph{deformers}. These specialized rigging tools are integrated into the production pipeline, affording artists a high degree of flexibility to convey their creative vision.

While physically based simulation can synthesize automatically rich elastic-like deformation on arbitrary shapes, physics models remain relatively slow and difficult to control in contrast.
Integration for physics remains orders of magnitude slower compared to procedural deformers, and this can impede real-time interaction needs.  Further, the handling of such simulations breaks the traditional production pipeline of iteration because the simulation must be computed as the result of the initial conditions and physical parameters, via numerical integration which restricts the ability to make interactive adjustments, and to control specific outputs at multiple points in the animation timeline. 

\vbz{In this work, we introduce an approach for adding expressive time-varying behavior to existing rigged character animation using directable deformers which create distinct purposeful effects such as springlike motion and time-delayed damping. }
Through parameterization we take advantage of the rigged skeleton and its animation to provide context and structure for the result. 
One exemplary deformer can generate an oscillation behavior on an articulated character, replicating a damped spring-like decay response similar to an elastic material. We also show a delayed deformation that creates a wavelike propagation along the skeleton of the character (Figure~\ref{fig:teaser}).  In both cases, these deformers are built from simple parametrized functionals with input from the animation and its derivatives, along with a small set of straightforward tuning values.
We avoid demanding the artist to set unintuitive physical parameters associated with a simulation, such as the Young's modulus of a physical material.  Our approach leverages the pre-existing animation and skeleton skinning weights from the hierarchy of the rigged surface
which provides cohesive kinematics information.  
When the skeleton is animated (with keyframes or simulation) the kinematics describes at the joint level the local body motion of the character's articulated skeleton, i.e. the joint translation or rotation. With the limb motions, dynamic skinning can be parameterized with position, angles, and velocity information, among others, which in turn can be used to drive our expressive deformers, which automatically blend and propagate along the rigged skin of the character.

Designing global space-time procedural deformation representing exaggerated elastic-like oscillations on an arbitrary 3D character is a difficult problem in general. The key insight of our approach is to split the deformation representation as a spatial-related one and a time-related one, while ensuring that deformation adapts automatically along the character hierarchy.
In this, Dynamic Skinning (DS) extends and generalizes the previous \emph{Velocity Skinning} (VS)~\cite{velocityskinning} and \emph{Acceleration Skinning} (AS) methods~\cite{accelerationskinning}, which achieve real-time spatial deformation on top of Linear Blend Skinning (LBS). Beyond the previous work, we generate a wider range of effects, including new \emph{time-based} deformers. Building upon the VS and AS frameworks, which address spatial deformation only, we propose an extended formulation %
with a comprehensive, time-dependent deformation using time-filtered joint
parameters. \vbz{Unlike the previous VS/AS work which supported no oscillation effects, DS creates } a more \emph{dynamic, time-varying} deformation effect for skeleton-based character animation.

Dynamic Skinning supports the following contributions.
First, we reformulate the VS/AS approaches, adding time-varying deformation which was not supported previously.
\dron{} \vbz{Second, we propose three general types of deformers, achieving respectively, an oscillatory behavior based on the joint velocity \vbz{(Section~\ref{sec:oscillation})}, a time delayed oscillation that accounts for the past joint velocities \vbz{(Section~\ref{sec:delayoscillation})}, and a delayed Linear Blend Skinning (dLBS) taking into account the past values of the joint pose \vbz{(Section~\ref{sec:dlbs})}.}\droff{} \vbz{Note, no oscillation effects appeared in previous work for AS/VS, greatly broadening the types of effects possible over this previous work.}
Third, we introduce novel controls to further customize deformer effects.  We showcase typical use cases, and comparisons of these effects in cartoon-like examples with state of the art.

\vbz{\section{Related Work}}

Much of the production animation created today for characters depends on skinning, such as Linear Blend Skinning (LBS)~\cite{magnenat1988joint}, where the surface geometry (skin) of the character is moved in relation to an articulated skeleton. The dependence between the joints of the skeleton and the vertices of the skin is defined by scalar \emph{skinning weights} and in combination with the skeleton and joints constitute
the \emph{rig} for character skinning.
LBS is a straightforward implementation of skinning where the deformation is computed as a linear combination of the joint transforms. A large set of work in computer animation research has improved upon LBS skinning techniques but LBS remains a common tool in practice.
As such, our general formulation for dynamic skinning is derived for the basic LBS formulation, but there is nothing that prevents the use of the proposed deformers to be applied to other methods, such as the \emph{Dual Quaternion} approach~\cite{kavan2008geometric} or other popular skinning techniques.

The described skinning techniques derive deformation of the character skin statically from the skeletal pose.  Aside from smooth skinning between joints, this approach offers no deviation between the skin and the skeleton to account for additional dynamic effects that may be desirable at the sub-bone level.  
To this end, physically-based simulation has been used to represent so-called \emph{secondary actions} on top of keyframe motion. We distinguish three main approaches to
compute secondary motion. 
First, projection-based dynamics~\cite{muller2005} and its extensions~\cite{bouaziz2014} have been proposed to model muscle and soft tissue movement effects, such as~\cite{Rumman_PPD_skinning_2014, iwamoto2015, komaritzan_i3d_2018}.  These can be tuned to exaggerate specific cartoon-like effects~\cite{Coros_SIGGRAPH_2012,Bai_SIGGRAPH_2016}.
Second, layering techniques have been proposed where distinct effects are handled as simple physical models, using jiggling implicit surfaces~\cite{Opalach_WorkshopAnimSimu_1994}, spring-driven bones~\cite{Kwon_CGF_2008}, or custom volume effects such as squash and stretch~\cite{angelidis_sca_2007}.
Third, reduced deformable models have been introduced~\cite{james_tog_2002,xu_siggraph_2016,RedMax2019} where precomputed shape analysis allows for model reduction, and therefore sped-up computation generating physically based response to primary motion.

\vbz{\subsection{Physical models}}

Recent techniques for producing simulation effects are still being proposed as well, such as~\cite{zhang2020,Benchekroun23, wu2023}.  We contrast our work to CODY~\cite{Benchekroun23} which treats the space of secondary effects as the orthogonal subspace of the rig.
This physically-based approach is able to model rich and detailed dynamic exaggerated deformations, possibly with collision handling, but CODY (and related work) has drawbacks. First, as physically based simulations, they are limited with respect to scale. Even fast approaches, that can tolerate models with thousands of vertices, cannot scale easily to a large number of elements in real-time --as may be required for video games applications-- or very detailed meshes of several millions of vertices.  In contrast, geometric methods (such as dynamic skinning) remain orders of magnitude faster. Second, simulation requires integration computed from initial conditions. This integration imposes that secondary animation be fully  simulated before being viewed, which hampers the efficiency of the animator and limits the possibility of navigating and editing details at arbitrary instances along the animation timeline. As such, standard production pipelines~\cite{ves_handbook, maya_deformer} often favor procedural deformers that are independent of past state when possible
(more on this subject in Section~\ref{sec:discussion}).
While several others have proposed reduced order models to approximate the dynamics with faster computation~\cite{hahn2013efficient,brandt2018hyperreducedpd,barbic2009,trusty2023subspacemfem}, the
speed of the simulation alone does not offer greater control over output animation.  In Barbi\v{c} et al.~\cite{barbic2009}, they aim to offer control through keyframes of the core motion, the approach and others related to it offer an indirect, non-linear relation between the inputs and the secondary motion of the outcome.  In contrast,
in addition to speed, dynamic skinning leads to more direct control over the output secondary effects that can be controlled easily and interactively.

\vbz{\subsection{Geometric models}}

Other geometric and kinematics-based deformations have been proposed to add toon-like effects for animation. Several techniques aim to define shape deformations for expressive effects from manual input~\cite{Li_SCA_2003, Kazi_CHI_2016,Dvoroznak_SIGGRAPH_2017,Ruhland_CS_2017} while other procedural approaches are more automatic, taking advantage of skeletal structure and kinematics.  
Noble and Tang~\cite{Noble_GRAPHITE_2006} introduce local bending parameterized by direction and velocity between consecutive bones. Kwon and Lee~\cite{Yong_TVCG_2012} propose \emph{squash and stretch} deformation parameterized by root joint velocity, and a \emph{drag} effect on end-effector joints for a skinned character.  
Kwon and Min~\cite{Kwon_LNEE_2012} proposed
to produce \emph{squash and stretch} effects
through bone stretching. 
Still other cartoon effects have been generated using time filtering~\cite{Wang_SIGGRAPH_2006} or oscillating splines~\cite{kass2008} applied to vertex trajectories. Recent related work~\cite{accelerationskinning} parallels these works in applying low-pass time filters in order to control the effect-specific timing and magnitude of effects relative to the instantaneous joint values of the skeleton.

In particular, the most similar to the ``Dynamic Skinning'' approach proposed in this paper are~\cite{velocityskinning,accelerationskinning} that introduce ``Velocity Skinning'' (VS) and ``Acceleration Skinning'' (AS), respectively.  The VS technique defines local deformers for \emph{squash and stretch} and \emph{drag} based on joint velocity and applies them as offsets to skinning positions. Meanwhile, as the name implies, the AS approach uses acceleration to add additional deformers such as \emph{followthrough} to the set.   
Like the work in this paper, these two previous works fit well within standard animation pipelines by reusing the existing rig (i.e. original skeleton and skinning weights) for ease in parameterization of deformation.  VS also introduces a closed form procedural deformation that can be computed extremely efficiently for each vertex in parallel, similar to basic skinning.
Our approach follows these, but extends the direction by generalizing the functional that influences the skinning, subsuming the other techniques into a single general framework.  In contrast, in dynamic skinning, we can direct a time-evolving response, such as a damped oscillation to create a spring and damper response or a propagation effect that moves along with time.  However, within our formulation, VS and AS are both reachable within the dynamic skinning framework, more on this in Section~\ref{sec:acceleration}.

\section{Dynamic Skinning Formulation}

We present the general formulation of Dynamic Skinning (DS) on a rigged character. The core idea of DS is to separate the space-time deformation applied to the surface of the mesh along simple controllable subparts: (c1) A spatial deformation parameterized by joint velocities that are at the origin of the character's motion; (c2) A spatial propagation of the deformation along the skeleton hierarchy and mapped to the surface, leveraging the existing rig; (c3) A dynamic time delay evolution obtained using a time filter applied to the joint velocities.

In the following, we first review in Sec.~\ref{sec:hierarchy-lbs-velocity} the velocity formulation for LBS expressed along the kinematic chain described by the hierarchy of the joints. This local hierarchical formulation provides an explicit relation between the velocity of the character's surface and the joint kinematic parameters that are at the origin of the joint transformations. 
Secondly, we define in Sec.~\ref{sec:ds-formulation} our hierarchical space-time deformers that take advantage of this parameterization to express the three levels of control described by (c1), (c2), and (c3).

\subsection{Hierarchy-based expression for LBS velocity}
\label{sec:hierarchy-lbs-velocity}

We formulate the expression of the vertex velocity in a LBS framework. \vbz{This \vbz{reiterates} the derivation initially introduced in VS~\cite{velocityskinning}. We include it here to provide a comprehensive formulation of the DS approach.}

Let us consider a character's surface defined by its mesh and associated animation skeleton that we assume to be represented as a single hierarchical tree structure. We call $\I$ the set of vertex indices, and $\J$ the set of joint indices of the skeleton. We denote $a(j)$ the ancestor (i.e. the parent) of the joint $j\in \J$ and we call $\mathcal{A}(j)$ the list of all the ancestors of joint $j$, from the root joint up to $j$ included. By convention, we will consider that the root joint is at index $0$, and its ancestor to be the global frame.

In the case of the so-called \emph{rigid skinning} (RS), a given vertex $i\in\I$, at initial position $p_i^0\in\mathbb{R}^3$ at rest pose, is rigidly bound to a single joint $j_0\in\J$. We call $p_{i/j_0}$ its absolute position after \emph{rigid skinning} deformation such that $p_{i/j_0}=T_{j_0}\,p_i^0$, with $T_{j_0}$ the rigid transform associated to joint $j_0$ expressed as a $4 \times 4$ matrix. An interesting representation of such a deformation is to consider vertex $i$ as being the extremity of a kinematic chain defined by $i$ and the set of joints $\mathcal{A}(j_0)$. As such, the net velocity of $i$, called $\dot{p}_{i/j_0}\in\mathbb{R}^3$, can be decomposed along the hierarchy of the kinematic chain as 
\begin{equation}
    \forall (i,j_0)\in(\I,\J)\,,\;\dot{p}_{i/j_0}=\sum_{j\in\mathcal{A}(j_0)} v_{i/j}\;\mbox{ ,}
    \label{eq:rigid_velocity}
\end{equation}
where $v_{i/j}$ is the velocity of $i$ induced by the transformation of joint $j$. Note that the velocity $v_{i/j}$ represents relative quantities with respect to a given joint expressed in the global reference frame. As each transformation associated to a single joint is a rigid transformation, $v_{i/j}$ can therefore be decomposed into a linear and angular velocity component such that 
\begin{equation}
    v_{i/j} = L_{j} + \Omega_j \times (p_i-q_j) \mbox{ .}
    \label{eq:linear-angular-velocity}
\end{equation}
$L_{j}$ is the linear velocity of joint $j$, and $\Omega_j$ is its angular velocity. $p_i$ is the current vertex position at index $i$, and $q_j$ is the position of the joint $j$ (see Fig.~\ref{fig:notation}-left). 

In \emph{Linear Blend Skinning} (LBS), the deformation transforming the vertex positions $p_i^0$ is expressed as a linear blending of the intermediate positions $p_{i/j}$ for varying values of $j$, or, in other words, can be interpreted as the linear blending of the previous kinematics chains. The deformed position $p_i^{\LBS}$ is thus
\begin{equation}
    \forall i\in\I\;,\;p_i^{\LBS} = \sum_{j\in\J} \alpha_{ij}\;p_{i/j}\;,
\end{equation}
where $\alpha_{ij}\in[0,1]$ are the LBS \emph{skinning weights}, and satisfy the barycentric property for each vertex $\sum_{j\in J} \alpha_{ij}=1$.

Differentiating this relation with respect to time and coupling it with Eq.~(\ref{eq:rigid_velocity}) leads to
\begin{equation}
    \forall i\in\I\;,\;\dot{p}_i^{\LBS} = \sum_{j\in\J} \alpha_{ij}\,\dot{p}_{i/j} = \sum_{j\in\J} \sum_{j_k\in\mathcal{A}(j)} \alpha_{ij}  \; v_{i/j_k}\;.
\end{equation}
While this formulation seems to involve a double summation, several local terms are summed multiple times when iterating over the hierarchy. Following the derivation described in VS~\cite{velocityskinning}, this relation can be rewritten as a simple summation representing a hierarchy-based expression for the LBS velocity
\begin{equation}
    \forall i\in\I\,,\;\; \dot{p}_i^{\LBS} = \sum_{j\in J} \beta_{ij}\; v_{i/j}\;,
    \label{eq:hierarchy-lbs-velocity}
\end{equation}
where the weights \(\beta_{ij}\), that we call \emph{hierarchical weights}, are a reformulation of the skinning weights \(\alpha_{ij}\) such that
\begin{equation}
    \beta_{ij}=\sum_{j_k\in\mathcal{D}(j)} \alpha_{ij_k}\,,
    \label{eq:hierarchical_weights}
\end{equation}
where $\mathcal{D}(j)$ is the set of all descendant joints of $j$ (including $j$ itself), see Appendix~\ref{sec:hierarchical_weights} for computational details. 
The \emph{hierarchical weights}, \(\beta_{ij}\), are related to the \emph{skinning weights}, \(\alpha_{ij}\), but have slightly different interpretation. While the \emph{skinning weights} represent the influence of a given joint $j$ over the global position of a vertex $i$, the \emph{hierarchical weights} describe how the local motion of the joint $j$ influences the vertex velocity. Like the \emph{skinning weights}, $\beta_{ij}\in[0,1]$. However, they do not satisfy barycentric constraints and are less sparse. 

We highlight that the formulation described by Eq.~(\ref{eq:hierarchy-lbs-velocity}) directly derives from standard LBS data but has several advantages. First, it provides an explicit relation between the final velocity applied to a vertex (i.e. $\dot{p}_i^{\LBS}$), and the contribution of each bone in the hierarchy that triggers it (i.e. $v_{i/j}$). Second, each contribution is automatically weighted according to its placement in the hierarchy thanks to the \emph{hierarchical weights}, $\beta_{ij}$.
Third, each weighted bone velocity is associated with rigid motion and can be decomposed into linear and angular components as with Eq.~(\ref{eq:linear-angular-velocity}), thus providing a precise description and parameterization of the motion associated with the bone-induced velocity component. 

\subsection{Hierarchical Space-Time Deformers}
\label{sec:ds-formulation}

\subsubsection{Velocity \vbz{and Acceleration} Skinning Formulation}
\label{eq:drag-effect}

The main idea of the original Velocity Skinning (VS) approach~\cite{velocityskinning} consists of leveraging hierarchy-based LBS velocity estimation to compute a per-vertex deformation on top of LBS  as the linear sum of procedural spatial deformers \(\psi\), 
parameterized by the vertex and the influencing joints, such as
\begin{equation}
    \displaystyle p_i^{\VS} = \underbrace{\sum_{j\in\J}\alpha_{ij}\; T_j\,p_i^0}_{p_i^{\LBS}} + \sum_{j\in J} \beta_{ij}\; \underbrace{\psi(\Omega_j,q_j,p_i^{\LBS})}_{\mbox{\scriptsize{deformer}}}, \\
    \label{eq:velocity-skinning}
\end{equation}
\dron{}
\vbz{
where $p_i^{\VS}\in\mathbb{R}^3$ is the VS deformed vertex position, and the procedural deformer $\psi$ is associated with a generic effect, typically modeled for translation, rotation, or scaling as a function of the joint angular velocity \(\Omega_j\), position \(q_j\), and current skinned vertex position $p_i^{\LBS}$. In other words, the \(\psi\) deformer converts the joint kinematic values into geometric deformations.
}
\droff{}
\vbz{Note, the VS formulation is limited to instantaneous effects that directly map the current time's joint velocity to a deformation.  Therefore, it relies on the joint velocity over time to produce a time-varying result.}

 \vbz{Acceleration Skinning (AS)~\cite{accelerationskinning} extended this model to allow \emph{followthrough} via the additional parameter from the joint acceleration \(\dot{\Omega}_j\). This was enabled by the fact that Eq.~(\ref{eq:hierarchy-lbs-velocity}) is valid for any higher-order derivative in time, including the second-order one for acceleration. However, AS is limited to animations with a single ``reversal'' effect driven from the acceleration term.  In particular, it does not produce longer, sustained (e.g., oscillating) effects beyond this single reversal.  These limitations of VS and AS are relaxed by our proposed time-varying advances. }

\subsubsection{Dynamic Skinning Formulation}

\begin{figure}[t]
    \begin{center}
      \includegraphics[width=0.9\linewidth]{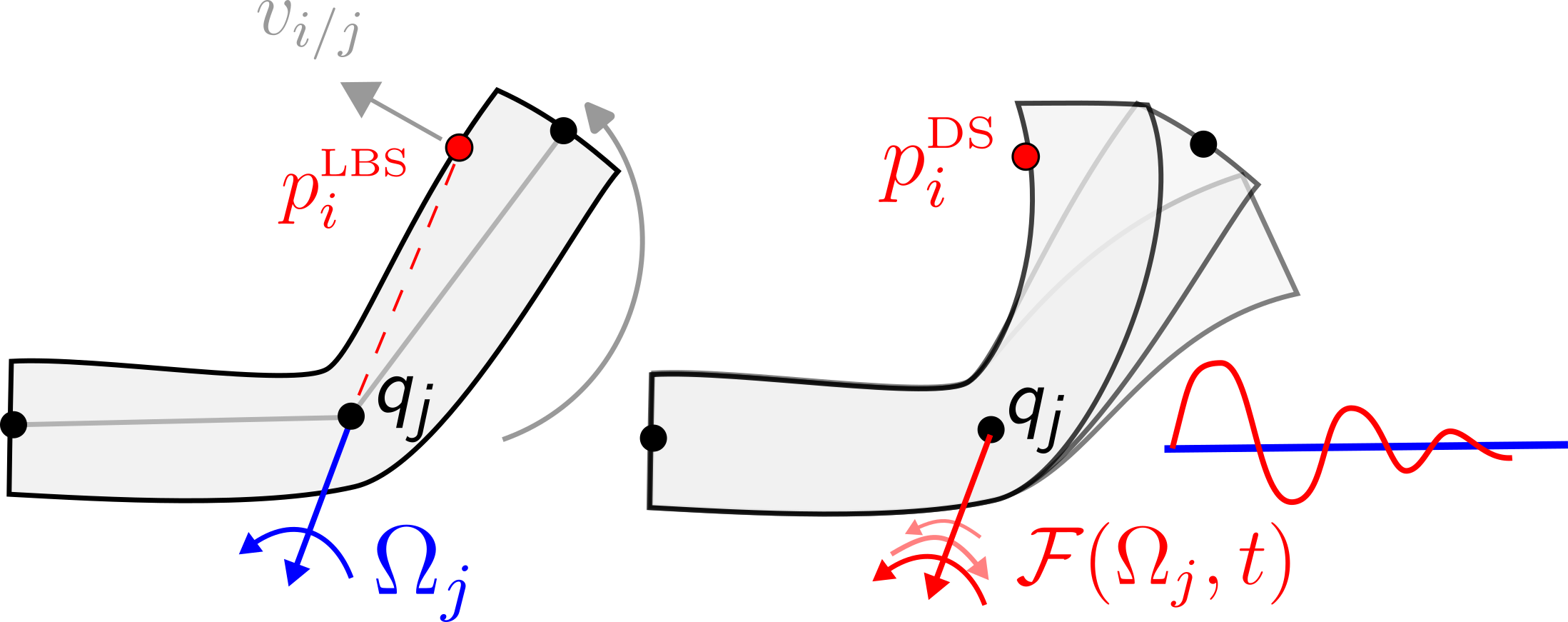}
      \caption{Left: A rotation applied to a cylinder using standard LBS along an angular velocity \(\Omega_j\) at joint $j$. Right: Dynamic oscillation obtained using a time filter \(\mathcal{F}(\Omega_j,t)\).}
      \label{fig:notation}
    \end{center}
 \end{figure}

We introduce the \emph{Dynamic Skinning} (DS) formulation to be more general and flexible than either VS or AS. Namely, our model integrates a \emph{time} dependence to the deformation, while keeping the existing formulation for the \emph{spatial} deformers $\psi$. 
To this end, we consider the velocity data of the joints to be functions of time: \(\Omega_j:t\to \Omega_j(t)\), where $t>0$ is the time variable, and interpret such kinematics data as an input signal that can be modified via time-filters \(\mathcal{F}\), before being used as parameter of the deformers. We propose the following expression for the general DS formulation of \emph{space-time} deformers: 
\begin{equation}
    \displaystyle p_i^{\DS}(t) = \displaystyle p_i^{\LBS}(t) + \sum_{j\in J} \beta_{ij}\; \psi\bigl(\mathcal{F}(\Omega_j,t),q_j,p_i^{\LBS}\bigr) , \\
    \label{eq:dynamic-skinning}
\end{equation}
where \(\mathcal{F}(\Omega_j,t)\in\mathbb{R}^3\) is the time-filtered value function of \(\Omega_j\) expressed at time $t$. 
\vbz{Thus, the DS method applies a spatial deformation on a modified time-filtered kinematic input described by the skeletal motion.  A representative comparison between VS and DS formulation is characterized in Fig.~\ref{fig:notation}.}
The new DS formulation provides simple control over the introduced criteria (c1), (c2), and (c3). Subpart (c1) representing the spatial deformation is encoded within the definition of the procedural function \(\psi\). Subpart (c2) is provided by the \emph{hierarchical weights} \(\beta_{ij}\) that derive from the pre-existing skinning weights. Finally, subpart (c3) is expressed via the time filter \(\mathcal{F}\), described in more detail in the next section.

\vbz{
In this paper, we specifically propose $\psi^{\txt{bend}}$, a procedural (time-varying) \emph{bending} deformer.  This deformer is a generalization of the \emph{drag} deformer originally introduced in VS \cite{velocityskinning}. This deformer, whose magnitude is controlled by a user-defined parameter $\gamma^{\txt{bend}}$, bends the shape in the direction opposite to the angular velocity by applying a rotation that increases with the distance to the joint center. We formulate it as
\begin{equation}
\psi^{\txt{bend}}(\Omega_j,q_j,p_i^{\LBS})=\Bigl(R^{\txt{bend}}\bigl(\Omega_j,p_i^{\LBS}-q_j\bigr)-I\Bigr)\;(p_i^{\LBS}-q_j)\mbox{ ,}
\label{eq:drag-deformer}
\end{equation}
where $R^{\txt{bend}}\bigl(\Omega_j,u\bigr)$ is a rotation matrix parameterized by its axis of rotation defined as $\Omega_j$, and angle defined as $-\gamma^{\txt{bend}}\,\|\Omega_j\times u\|$, \(u\in\mathbb{R}^3\) is the vector between the point and the center of rotation, and $I$ is the identity matrix, so that $\psi^{\txt{bend}}$ is a displacement that vanishes when $\Omega_j=0$.  
}

\vbz{
While we exclusively utilize the bend deformer in this paper, to specifically target springlike response, there is no loss of generality in the application of DS in other deformers from VS or AS.  Further, 
we contrast the bend deformer to that of the drag one from VS in that bending does not act as an instantaneous drag response, but as a general bending function that is capable of a spectrum of time-varying effects. One key example is adding oscillation effects, described next.
}

\section{Modeling Dynamic Oscillation Effects}
\label{sec:oscillation}

In this section, we present the general formulation for the time filter and the use of the different kinematics parameters of the joints to achieve dynamic oscillation compatible with \emph{secondary action effect}.
Defining a time filter in general can be a hard problem; in our case, we propose to consider $\mathcal{F}^{\osc}$ to be a linear causal filter on a limited time-window \(W\) to remain efficient to compute and simple to define.
We aim to represent an oscillation effect that typically takes place after a sudden change of velocity, while being smoothly damped when the velocity is stopped or remains constant. We propose to design this behavior via the definition of the response of the filter to a unit step function \(U(t)\), i.e. an input signal in velocity such that \(U(t)=0\) for \(t<0\) and \(U(t)=1\) for \(t\geq 0\). Calling $h(t)$ the response of the filter to this unit step, we propose the following representation
\begin{equation}
    \begin{array}{ll}
    h(t) & = \left(1+\kappa(t)\right)\,U(t)\;, \\
    & \mbox{ with }
    \left\{
        \begin{array}{l}
            \kappa(0)=0 \\
            \kappa(W)\simeq 0 \\
            \max_t |\kappa(t)| \leq A
        \end{array}
    \right.
    \end{array}
    \label{eq:kernel_properties}
\end{equation}
The value one enforces the filtered response to track the input signal, while \(\kappa\) is a user-defined function that we call \emph{oscillating kernel}, acts as a temporary perturbation.

In the default case, we will consider a typical \emph{oscillation kernel} to be defined as
\begin{equation}
    \kappa(t)=A\,\exp(-t/\tau)\,\sin(2 \pi f\,t)\;,
    \label{eq:kernel_sinus}
\end{equation}
representing an oscillation with a frequency \(f\) with a magnitude bounded by \(A\) for \(t\geq 0\) which is exponentially damped with a characteristic time \(\tau\). Considering \(W>5 \tau\) for instance satisfies the condition that \(\kappa(W)\) is negligible. 

The general expression of the filtered signal for an arbitrary time-varying input \(\Omega_j\) and \emph{oscillation kernel} can be computed as the convolution between the input angular velocity signal and the so-called \emph{impulse response}, which is the derivative (in the sense of the distribution) of its response to a unit step. 
We note the impulse response \(\dot{h}\), and we have
\begin{equation}
\dot{h}(t)=\delta(t)+\dot{\kappa}(t)\,U(t)\mbox{ ,}
\end{equation}
where \(\delta(t)\) is the Dirac distribution. Then, the general expression of the filtered signal can be written as

\begin{equation}
    \begin{array}{ll}
    \mathcal{F}^{\osc}(\Omega_j,t) & = (\dot{h}\;\star\; \Omega_j)(t) \\
     & \displaystyle =\Omega_j(t) + \int_{u=0}^{W}\dot{\kappa}(u)\,\Omega_j(t-u)\;\mathrm{d}u\,.
    \end{array}
    \label{eq:convolution}
\end{equation}
where \(\star\) is used to express the convolution, and \(\dot{\kappa}\) is the time derivative of the oscillation kernel.
Finally, an alternative formulation can be obtained by applying an integration by parts to the integral, which leads, as shown in Appendix~\ref{appendix:integration}, to
\begin{equation}
    \begin{array}{l}
    \mathcal{F}^{\osc}(\Omega_j,t) = \displaystyle \Omega_j(t) + \int_{u=0}^{W} \kappa(u)\;\,\dot{\Omega}_j(t-u)\;\mathrm{d}u\,,
    \end{array}
    \label{eq:convolution2}
\end{equation}
where \(\dot{\Omega}_j\) is the time derivative of the angular velocity, i.e. the angular acceleration.

The main advantage of our formulation is its ability to handle a user-defined \emph{oscillating kernel}. The default kernel proposed in Eq.~(\ref{eq:kernel_sinus}) provides a simple way to control the notion of frequency, magnitude and damping fully independently, and complements the spatial deformation related to the bend effect whose magnitude is parameterized by \(\gamma^{\txt{bend}}\). More generally, the function of the oscillation kernel can be fully tuned with different representation to provide additional degrees of freedom, or effects, as discussed in the results in Sec.~\ref{sec:kernel}.

Our numerical implementation of the convolution from Eq.~(\ref{eq:convolution2}) is obtained from simple summation and finite difference. Assuming \(\Delta t\) is a constant time step, and calling the finite time \(t=i\,\Delta t\), the numerical filtered angular velocity is computed as
\begin{equation}
    \begin{array}{ll}
    \mathcal{F}(\Omega_j,i \Delta t)=\Omega_j(i \Delta t) \; + 
    \\ \displaystyle 
    \sum_{j=0}^{j<W/\Delta t} \kappa(j\Delta t)\;\,\bigl(\Omega_j((i-j)\Delta t)-\Omega_j( (i-j-1)\Delta t)\bigr)\,.
    \end{array}
    \label{eq:convolution_numerical}
\end{equation}

Our approach contrasts with the more classical use of mass-spring-damper system that is typically modeled by a second-order recurrent relation.
While such a system is numerically simpler to express, it offers fewer degrees of freedom to the user. In particular, the frequency of the oscillation in a mass-spring system is coupled with its damping, while we propose a method able to tune these parameters fully independently. Additionally, our approach does not suffer from stability concerns, as it does not require iteration over previous values which ensures that the results are unconditionally stable for any chosen parameters, even for low damping values and high frequency.

\dron{}\vbz{
In Appendix \ref{appendix:pseudo-code-DS}, we provide a comprehensive pseudo-code of the complete DS algorithm, emphasizing its computational simplicity: it consists merely of basic loops over the vertices, joints, and past angular velocities coupled with user-defined procedural functions, making it fully compatible with GPU implementation.}
\droff{}

\section{Time Delay Filter}
\label{sec:timedelay}

The oscillation effect we proposed in the previous section depends on the current and past state of the angular velocity of the joints, however every vertex of the mesh is influenced in parallel by the same time values. In this section, we propose an additional time-filtering effect, modeled as an explicit per-vertex time delay, able to represent a deformation propagating in time throughout the character similar to an advancing wave.
We propose the use of such time delay to generate two additional effects. First, a time delayed oscillation, as a direct extension of the previously introduced oscillation framework. Second, a delayed Linear Blend Skinning (dLBS) that applies a time delay directly to the skinning computation itself. We then show that the two effects can be combined to generate a user-controlled deformation propagation along the skeleton hierarchy.

\subsection{Formulation of the delay filter}
\label{sec:delay-filter}

Let us first introduce the general time delay filter \(\mathcal{F}^{\delay}\) as a pure delay applied to an arbitrary time-varying signal \(x\) such that
\begin{equation}
    \mathcal{F}^{\delay/i}(x,t) = x(t-d_{ij}/c)\;,
\end{equation}
where \(d_{ij}\) is a representation of a distance between a vertex $i$ and a joint $j$, and \(c\) is the velocity of the propagation along the character.

\begin{figure}[htb]
    \centering
    \centerline{\includegraphics[width=0.65\linewidth]{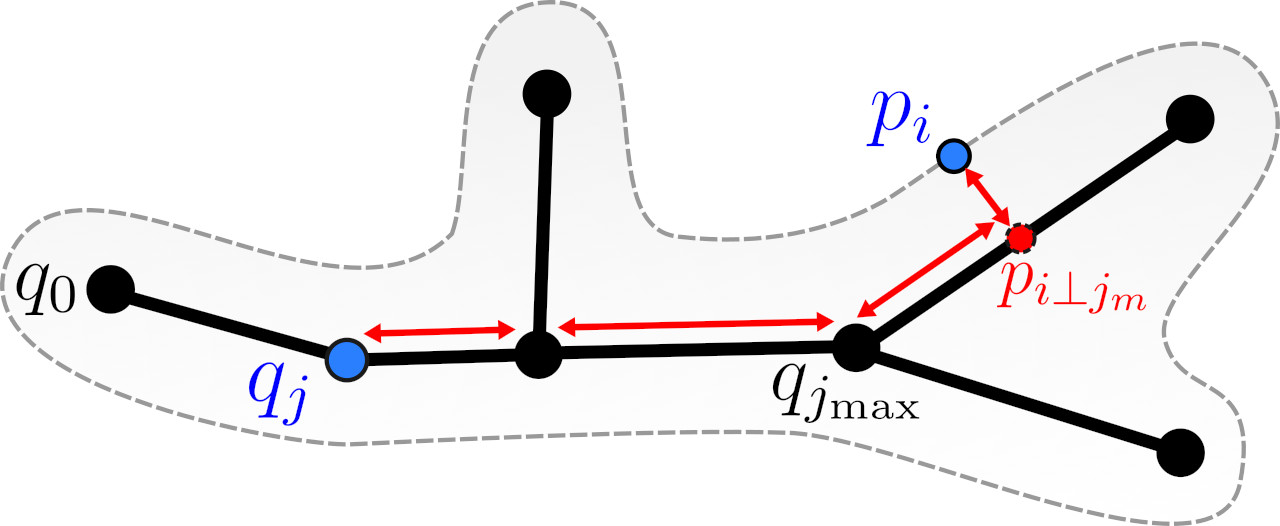}}
    \caption{$d_{ij}$ from Eq.~(\ref{eq:distance}) is the distance between the vertex $i$ and joint $j$ along the skeleton, corresponding to the length along the red arrows.} 
    \label{fig:distance-delay}
\end{figure}

To give an appearance of deformation propagation along the hierarchy of the character, we compute \(d_{ij}\) composed of two contributions. First, the closest distance between the vertex \(i\) and its closest position on its most influencing bone $j_{\max}$ such that \(j_{\max}=\mbox{arg}\max_j(\alpha_{ij})\). Second, the distance along the skeleton itself up to the joint \(j\), computed as the cumulated lengths of the bones along the ascending hierarchy of the skeleton up to $j$, assuming $j\in\mathcal{A}(j_{\max})$.
\begin{equation}
    \begin{array}{ll}
        d_{ij} = & \|p_i-p_{i \perp j_m}\|+\|p_{i \perp j_m}-q_{j_{\max}}\| \\ & + \displaystyle \sum_{j'\in\mathcal{A}(j_{\max})}^{a(j')=j} \|q_{j'}-q_{a(j')}\|.
    \end{array}
    \label{eq:distance}
\end{equation}
The decomposition of such distance is illustrated in Fig.~\ref{fig:distance-delay}, and all distances between each vertex and joint can be precomputed in the rest-pose of the character.

\subsection{Effect: Delayed oscillation}
\label{sec:delayoscillation}

The first effect we propose is to apply the time delay filter as an additional layer on top of the oscillation effect.
To this end, we define the global time filter \(\mathcal{F}\) from Eq.~(\ref{eq:dynamic-skinning}) as the composition of the oscillation and the delay filter as
\begin{equation}
    \mathcal{F}(\Omega_j,t)=\mathcal{F}^{\osc}\circ \mathcal{F}^{\delay/i}(\Omega_j,t)\;.
\end{equation}
where \(\mathcal{F}^{\osc}\) is the oscillation filter from Eq.~(\ref{eq:convolution2}).
This results in an oscillating bend effect that starts with a delay, and propagates along the skeleton hierarchy with a speed \(c\) after the joint starts moving.

\subsection{Effect: Delayed Linear Blend Skinning - dLBS}
\label{sec:dlbs}

We propose a second effect related to the time delay filter by applying it on top of the Linear Blend Skinning position computation itself, extending this time the way we compute LBS, rather than the bend-like deformation as previously described. We then introduce a \emph{delayed LBS} (dLBS) as
\begin{equation}
p_i^{\dLBS} = \sum_{j\in\J} \alpha_{ij}\; T_{j}^{\delay/i}(t)\;\,p_i^0\;.
\end{equation}
\(T_{j}^{\delay/i}(t)\) represents the global transformation applied to the vertex $i$ considering the delay applied to the hierarchy of the joints. The explicit computation of this delayed transformation can be obtained by the product of all local transforms $T_{j'}^{\local}$ for $j'\in\mathcal{A}(j)$ as
\begin{equation}
    \begin{array}{l}
        T^{\delay/i}_{j}(t) \displaystyle =\prod_{j'\in\mathcal{A}(j)} T_{j'}^{\local}\left(t-\frac{d_{ij'}}{c}\right) \\
        \displaystyle  = T_0^{\local}\left(t-\frac{d_{i0}}{c}\right)\; \cdots \; T_{a(j)}^{\local}\left(t-\frac{d_{ia(j)}}{c}\right) T_{j}^{\local}\left(t-\frac{d_{ij}}{c}\right)
    \end{array}
\end{equation}

\begin{figure}[]
    \centering
    \centerline{\includegraphics[width=\linewidth]{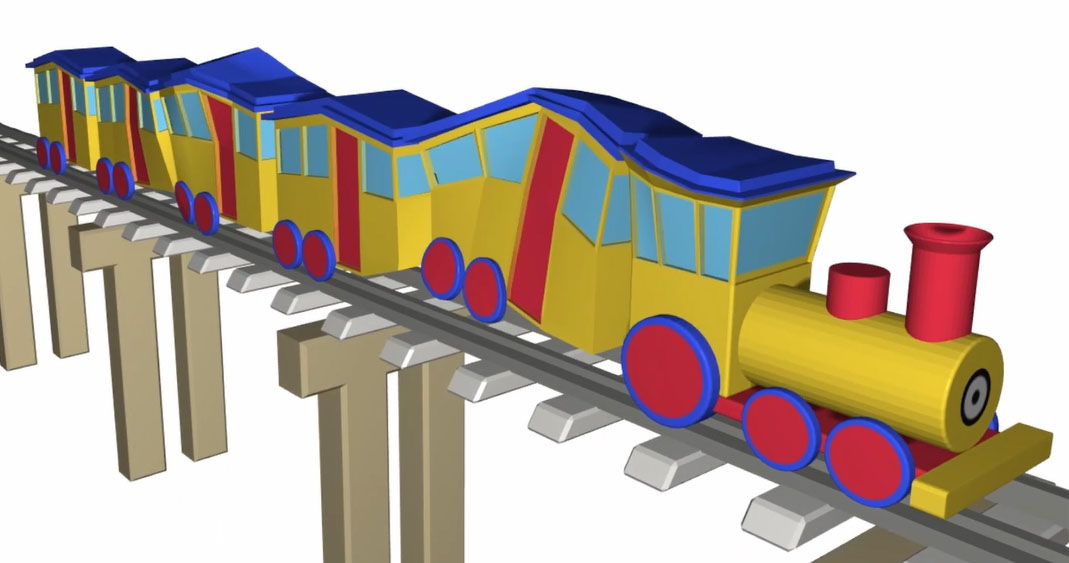}}
    \caption{The toy train showcases the effect we call ``delayed'' Linear Blend Skinning (dLBS) which propagates oscillation effects at the engine along the skeleton rig to the caboose.}
    \label{fig:train}
\end{figure}

Finally, a delayed deformation can be applied within our dynamic skinning framework by replacing \(p_i^{\LBS}\) by \(p_i^{\dLBS}\) in Eq.~(\ref{eq:dynamic-skinning}). This enables the representation of a delayed, and therefore apparently propagating, skinning articulation (Figure~\ref{fig:train}).

The combination of the two effects of delayed oscillation and the dLBS offers a more integrated appearance in the final animation where both the oscillation and the skeletal effect reinforce the time propagation along the skinned character.

\section{Controllability}

Basic controllability for DS is straightforward, for example for the oscillation effect control is of the direct amplitude, damping and frequency of the desired wiggle - which are all parameters that animators understand well.  Further, the effects can be layered over existing animations through the described straightforward tuning - which subsequently results in expected augmentations to the input skeletal animation due to the effect.  In contrast, in comparable secondary effects that are physically based, nonlinear response changes make tuning parameters of material stiffness and damping less intuitive and unwieldy, especially when trying to generate complex or layered  effects.  Beyond basic controls, here, we describe a number of helpful advanced control capabilities for DS effects.

\begin{figure}[htb]
    \centering
    \centerline{\includegraphics[width=\linewidth]{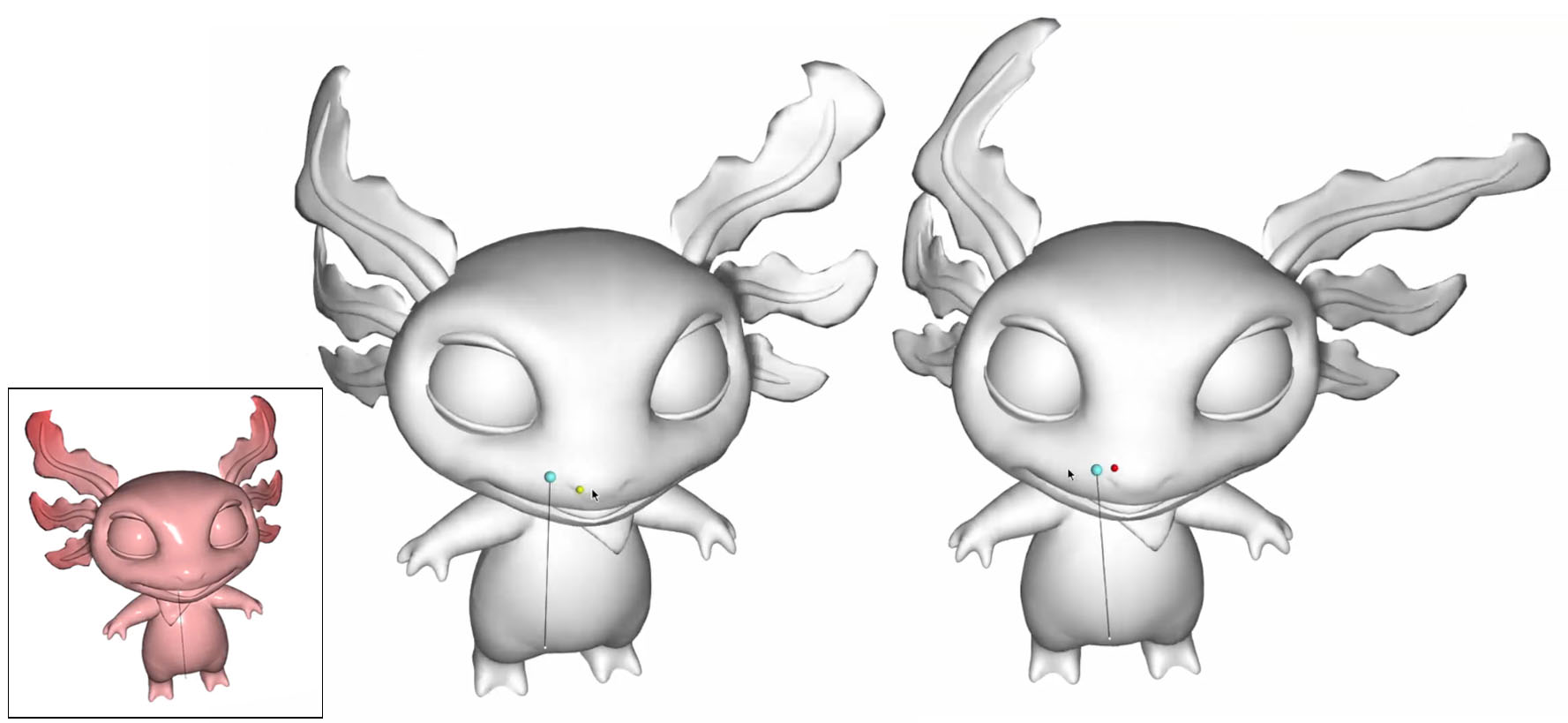}}
    \caption{Frames from an example of artist-driven vertex weight painting (inset) to add exaggerated effects on the extremities of the head of the character.} 
    \label{fig:vertex_weights}
\end{figure}

\subsection{Weighted deformers}
\label{sec:weight_painting}

We explore two techniques for refining control of the influence for deformers through weights.  
\dron{}\vbz{
A simple approach, as described in previous work in velocity skinning, consists in setting the value \(\gamma^{\txt{bend}}\) to be a variable function of the vertex index, i.e. \(\gamma_i^{\txt{bend}}\) for index $i$. Such vertex weight painting can be used to give fine-grained control over where the effect is visible. 
The benefit of this is that it gives an animator full ability to tune an effect based on features that may be present in the model but are not distinct in the skeleton. For example, the head extrusions on the stylized character in Figure~\ref{fig:vertex_weights} have no bones but vertex weighting can bring them to life with an oscillatory response when the head moves.
Such per-vertex control can be seamlessly extended to the parameters of the oscillating kernels, such as the magnitude $A$, damping $\tau$, and frequency $f$ from Eq.~(\ref{eq:kernel_sinus}).
Additionally, these parameters can further have a dependence on the joint index, the original skinning weights can smoothly transition these weights to derive a vertex-level mapping of a deformer, this concept works well when the features that an animator seeks to tune align with the skeletal bones.  For example, to make the influence vary along a limb, or other extremity.  We used variants of these two weighting techniques in the accompanying video in results for the clown model, to animate the hat and arms for the clown, respectively.}
\droff{}

\subsection{Oscillation kernels}
\label{sec:kernel}

\begin{figure}[htb]
    \centering
    \centerline{\includegraphics[width=\linewidth]{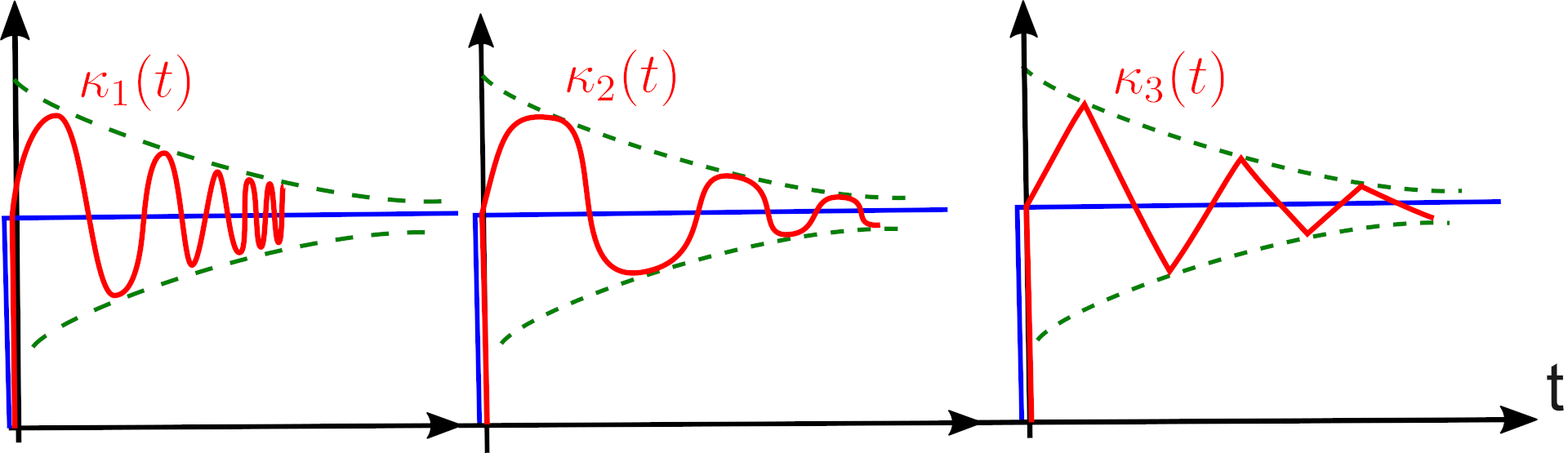}}
    \caption{Examples of custom oscillation kernels leading to various dynamic oscillation effects.} 
    \label{fig:oscillation_type}
\end{figure}

We proposed in Eq.~(\ref{eq:kernel_sinus}) the use of a simple oscillating kernel featuring a sinusoid with a constant frequency. 
Beyond, our formulation can handle different types of oscillation kernels as illustrated in Fig.~\ref{fig:oscillation_type} as long as they satisfy the properties defined in Eq.~(\ref{eq:kernel_properties}), and can possibly be provided either as parametric curves, or directly as a procedural function that could be defined in a shader.  

\noindent Appendix~\ref{appendix:kernel} describes the details that build the kernels shown.

\subsection{Deformation origin control}
\label{sec:deformation-origin-control}

An additional customization to the dynamic skinning framework is the ability to adjust the center of rotation associated to the deformation \(\psi\) as well as its angular magnitude.
In the default formulation proposed in Eq.~(\ref{eq:drag-deformer}), the bend deformation is applied as a rotation centered around the joint at the origin of the motion, i.e. joint associated with a non-zero angular velocity. However, this behavior can be adjusted for artistic purposes. 
\begin{figure}[tb]
    \centering
    \centerline{\includegraphics[width=\linewidth]{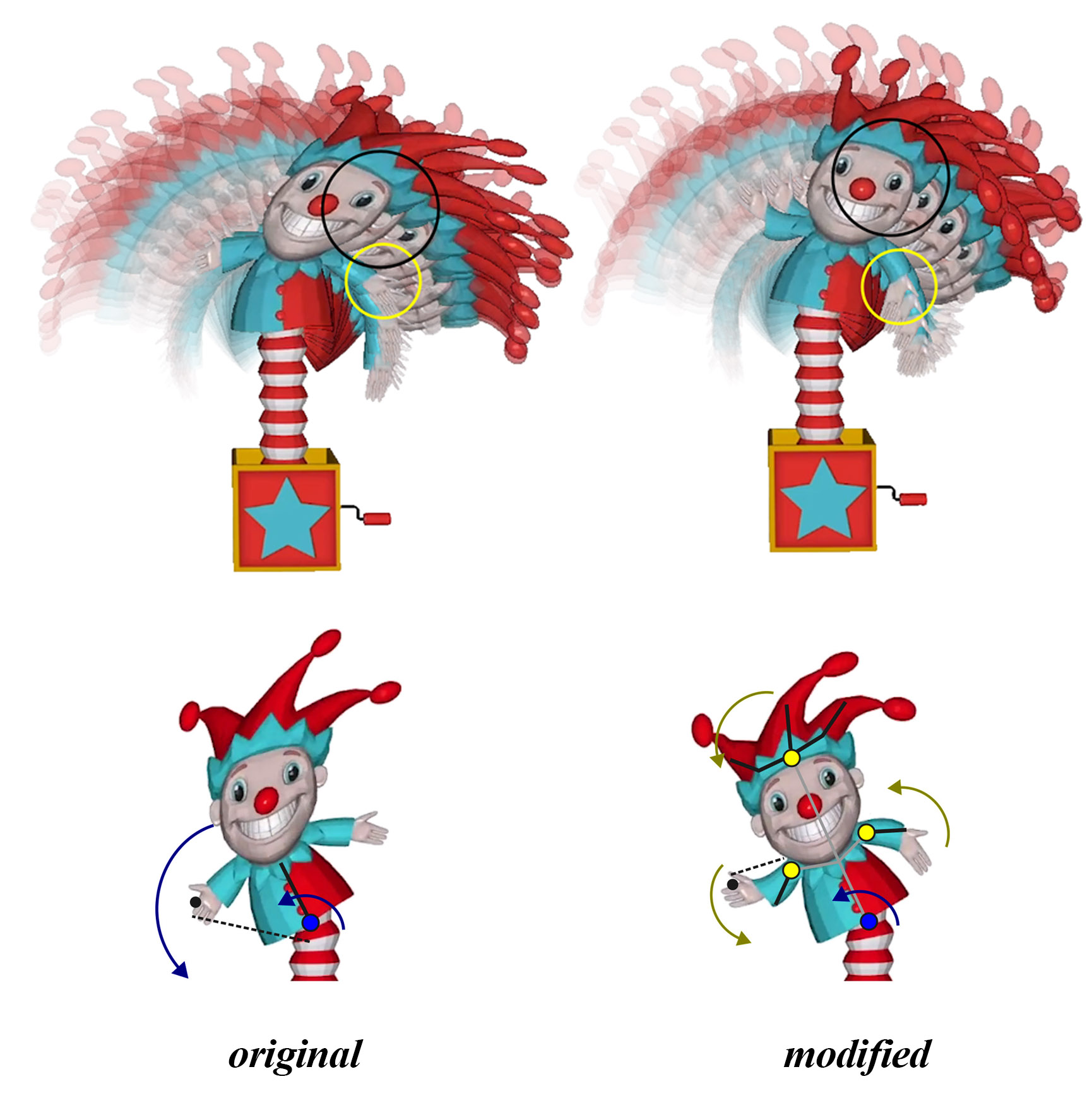}}
    
    \caption{Comparison of standard bend effect (left) with isolated deformation of the hat and hands (right) accomplished by changes in centers of rotation.  Namely, right uses local centers of rotation at the start of each extremity (yellow dots).
    Top shows results from center rotation control.  Black circle highlights the effect on the near rigid head, while the yellow circle shows the impact of the deformation on the hands. \vbz{While both effects may have merit in distinct cases, the right offers a new context specific manner in which the artist may guide the outcome of the effect based on the model's features.}} 
    \label{fig:bone_center}
\end{figure}

A typical application of this extension is to model shapes with a rigid base and flexible extremities using the existing rig. The fundamental idea is that when rotation occurs within the rigid section, the deformation of the flexible extremity is computed as if it originates from the joint marking the beginning of the flexible section.
Let us consider a joint \(j_{\min}\) be such a joint, and \(O_{j_{\txt{min}}}\) the \emph{flexible origin} operator such that 
\begin{equation}
    O_{j_{\txt{min}}}(j)=
    \left\{
    \begin{array}{ll}
        j_{\min} &\mbox{ if } j_{\min}\in\mathcal{D}(j) \mbox{ ,} \\
        j &\mbox{ otherwise.}
    \end{array}
    \right.
\end{equation}
We consider the following bend deformer variation with customized origin:
\begin{equation}
    \begin{array}{lrl}

    \psi_{s}^{\bend}=& R_s^{\bend}(\Omega_j,p_i^{\LBS}-q_{O_{j_{\txt{min}}}(j)})&(p_i^{\LBS}-q_{O_{j_{\txt{min}}}(j)})\mbox{ .} \\
    \end{array}
\end{equation}
An illustration of the effects is shown in Fig.~\ref{fig:bone_center}, where the clown model is animated with a rotation applied on the joint at the base of its center-body. Fig.~\ref{fig:bone_center} (left) shows the default effect where the whole model is flexible, while the right shows the customized case where the main body is rigid, while the extremities (each arm and the hat) are flexible. 
Each extremity has its own rotation center (yellow dots) at the beginning of their limbs. The effect accentuates the visual feedback of each flexible extremity as if they were independent of each other.

This control can be seen as complementary to the weight painting presented in Sec.~\ref{sec:weight_painting}. Weight painting allows a per-vertex control, but requires the user to make sure that the weights are smoothly varying across the model, and only acts at the level of the magnitude of the deformation. In contrast, this custom origin control relies on the skinning rig and ensures a smooth transition between the rigid and the flexible parts, but requires a skeleton in the targeted limbs. Furthermore, this extension not only changes the magnitude of the deformation but also the shape of the deformation itself by changing the center of the rotation as can be seen in the figure.

\section{Results and Extensions}

We showcase a variety of results in the accompanying video that captures a snapshot of the effects that are possible with our approach.  
Indeed, the scope of dynamic skinning is broad and we have not exhausted the number of different ways the framework can be used.  Here we describe several \vbz{ablations/experiments that highlight effects for: 1) adding oscillation with comparisons of the effect's parameters (amplitude, damping, etc.) and different kernels~(Section~\ref{sec:kernel}); 2) dLBS that highlights how it impacts especially long chains of bones, such as the train, in the accompanying video; 3) we show the combined effect of the two; and 4) we highlight the impact of weighting parameters and control techniques with various outcomes~(Section~\ref{sec:weight_painting}).  Further, we make qualitative and quantitative comparisons between our own and previous results. }
Note, all animations presented in the accompanying video are obtained from real-time capture using either a CPU implementation for the low resolution model, or a GPU implementation computed in a single pass vertex shader. 

We compare our technique to the recent work, CODY~\cite{Benchekroun23}, which offers a fast computation for secondary motion by physical simulation of the complementary dynamics of the input (keyframe) motion.  Quantitatively, their tool is very fast in terms of physical simulation, but ours is significantly faster (see timing results below) because we do not have to compute a physical simulation and our implementation can exploit significant speed ups using a GPU.  \vbz{ In fact, our implementation is orders of magnitude faster than the state of the art.}

Qualitatively, \vbz{and as summarized in Appendix~\ref{appendix:parameters}, presenting an overview of the user-defined parameters,} we also believe our tool is more amenable to artists' needs as
\vbz{it is very straightforward to use, with only a few intuitive parameters yielding an expected and high quality result. In addition, }
it is very customizable, down to the vertex level control through weight painting.  In contrast, CODY enhances the character motion with physical attributes automatically, but offers \vbz{less direct and less intuitive control over the outcome.  As with other physical simulation tools, the tuning of CODY results can be cumbersome, as it happens through specification of control for material and physical properties as well as initial conditions, which impact the resulting simulation}.  An accompanying video of a manta ray shows the two techniques applied to the same base motion and model.

\begin{table}[!h]

\begin{center}
\caption{This table highlights the relative computational cost (computation time per frame normalized by the number of polygons) and the largest size of models used in comparable state-of-the-art.}
\begin{tabular}{c  c  c } 
 \hline
   & Timing/\#polygon $\downarrow$ & Model size (\#polys) $\uparrow$ \\ 
[0.5ex] 
 \hline
 Ours & \textbf{0.00003 ms/polygon} & \textbf{1.5M}  \\ 
 Wu et al.~\cite{wu2023} & 0.004 ms/polygon & 4.2K \\
 CODY~\cite{Benchekroun23} & 0.05 ms/polygon & 0.3M  \\

 \hline
\end{tabular}
\end{center}
\end{table}

\noindent The timing results above compare the computational performance of our work to CODY~\cite{Benchekroun23} and Wu et al.~\cite{wu2023}.  
As a simulation tool, CODY appears much slower than the others, while the approach of Wu and colleagues is faster. Neither shares the range of model sizes we show in our GPU implementation.

\subsection{Acceleration skinning formulation}

\label{sec:acceleration}

The method of \textit{Acceleration Skinning} (AS)~\cite{accelerationskinning} was introduced to extend the drag-like deformation to the notion of \textit{followthrough}. The method relies on utilizing the additional input of the angular acceleration \(\dot{\Omega}_j\), and uses it to separate the acceleration phase from the deceleration phase. We show that DS provides a more global framework in which similar effects can be replicated.  In addition, due to the formulation ours is more easily tuned over the original approach.

We propose to replicate AS in the proposed DS formulation by extracting the angular acceleration as a simple resultant of a derivative filter \(\dot{\Omega}_j(t)=1/(2\,\Delta t)\,\left(\delta_{\Delta t}-\delta_{-\Delta t}\right)\star \Omega_{j}(t)\). Then, the deformation \(\psi\) can be adapted to match the one proposed in AS, as
\begin{equation}
    \left\{
    \begin{array}{l}
    \psi^{AS}=a \,\psi(\Omega_j(t),q_j,p_i^{\LBS}) + (1-a) \,\psi(\dot{\Omega}_j(t),q_j,p_i^{\LBS})\mbox{ ,} \\
    a = \mbox{clamped}(\dot{\Omega}_j(t)\cdot \Omega_j(t),-1,1).
    \end{array}
    \right.
\end{equation}
where the first part corresponds to the drag effect during the acceleration phase, and the second part is the followthrough effect applied during the deceleration phase.%
More generally, representing the bend deformer for oscillation using two main lobes (one positive, one negative), will also allow for the modeling of the drag, and followthrough effect. Using the oscillation kernels will further add the possibility to fine-tune the function as explained in the previous section, and avoids the direct use of the angular acceleration which can suffer from numerical instability.

\subsection{Computational speedup}

Eq.~(\ref{eq:convolution2}) provides a general way to compute an oscillating-like behavior in response to an arbitrary input signal and can be computed efficiently in a shader for real-time results. Still, the full computation of the convolution requires, at each frame, the weighted sum over the previous values of the angular velocity, and therefore remains more costly than a direct computation of an instantaneous drag effect from VS~\cite{velocityskinning}. 

\begin{figure}[htb]
    \centering
    \centerline{\includegraphics[width=1.0\linewidth]{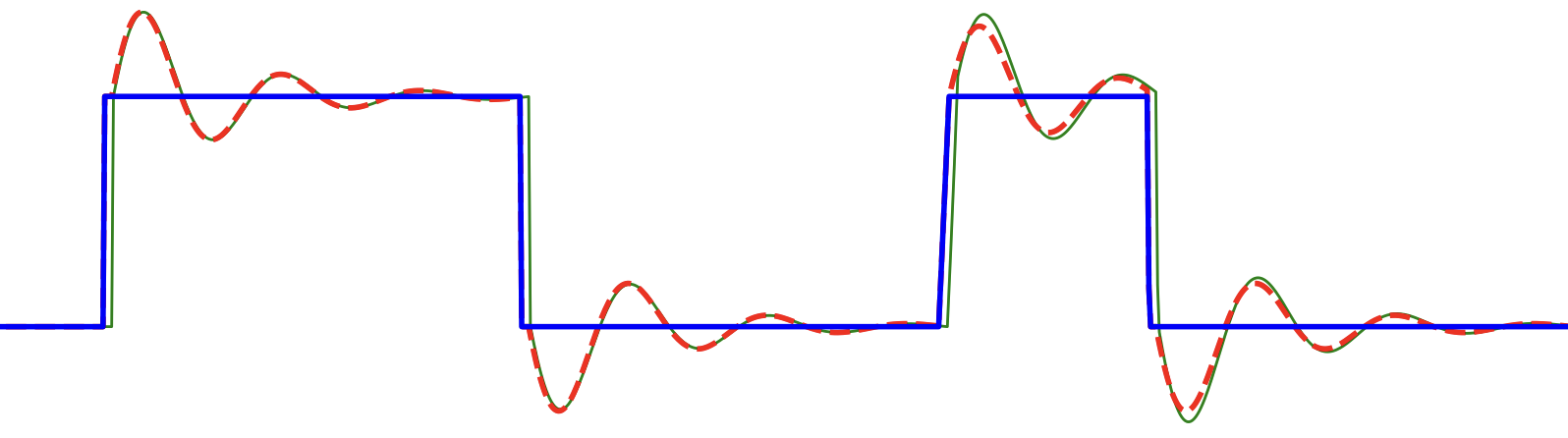}}
    \caption{Oscillations obtained on a step-like signal (blue curve) when using the full convolution from Eq.~(\ref{eq:convolution2}) (green curve), and the accelerated version from Eq.~(\ref{eq:acceleration}) (red dashed).} 
    \label{fig:acceleration}
\end{figure}

We propose an efficient alternative approximation in the case where the animation exhibits sudden impulse-like motions that do not require such storage and weighted summation while still preserving the appearance of the oscillation effect. To this end, we consider only the last variation of angular velocity leading to the most significant contribution to a single oscillation.
We then compute our approximated signal as
\begin{equation}
    \mathcal{F}(\Omega_j,t) \simeq \Omega_j(t)+\kappa(t-\hat{t}) \; \dot{\Omega}(\hat{t}) ,
    \label{eq:acceleration}
\end{equation}
where \(\hat{t}\) is the time of the maximal contribution updated at each frame as
\begin{equation}
    \hat{t}=
    \left\{
    \begin{array}{ll}
        t &\mbox{ if } \;\; \Omega(t)>e^{-(t-\hat{t})/\tau}\,\Omega(\hat{t}) \mbox{ ,} \\
        \hat{t} &\mbox{ otherwise.}
    \end{array}
    \right.
\end{equation}
As shown in Fig.~\ref{fig:acceleration}, the oscillation behaviors obtained in the case of impulse-based motions with the full convolution and with our accelerated version are very similar.

\begin{figure}[htb]
    \centering
    \centerline{\includegraphics[width=\linewidth]{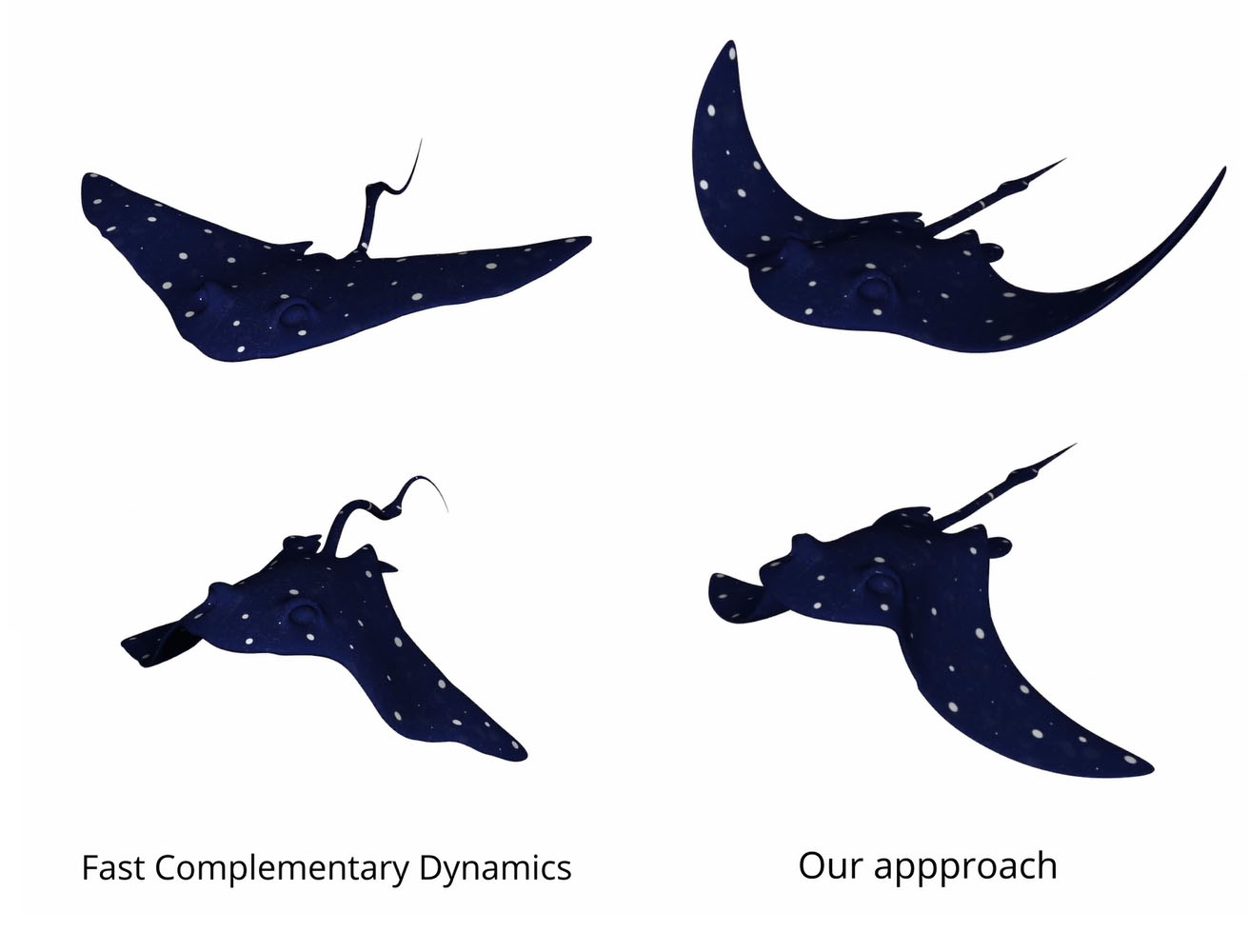}}
    \caption{\vbz{Comparison with Complementary Dynamics (CODY).  CODY yields full body effects that are difficult to control because the simulated results are tuned indirectly through physical parameters which do not lead to intuitive changes in results.  For example, the tail of the CODY manta shown here moves on its own when the wings are tuned - with a specific wavelike response in the tail that changes significantly with whole-body damping and stiffness.  In contrast, in our approach the tail's motion can be suppressed by vertex painting, while the wings move smoothly based on intuitive amplitude guiding values in the deformer.  Likewise the tail can be built to have its own desired effect with the same intuitive parameters if the artist wishes.}} 
    \label{fig:ray_compare}
\end{figure}

The key advantage of this approximation is to convert the full computation of the convolution described in Eq.~(\ref{eq:convolution_numerical}) requiring the weighted summation over a few hundred samples (\(W/\Delta t\)) at each frame per joint, to a single computation and comparison per joint. The computational complexity of this acceleration is therefore the same as Velocity Skinning. This approximation is particularly effective when the joint motion is made of a succession of sudden impulses, but does not integrate the full history of the past angular velocity if a more continuous motion change is performed.

\section{Discussion and Conclusion}
\label{sec:discussion}

An important benefit of dynamic skinning and related techniques is the way in which the work dovetails with other efforts in production.  Contrasting the efficiency driven efforts of producing real-time physically plausible secondary motion using CODY/complementary dynamics~\cite{Benchekroun23}, our technology supports an artist-driven workflow by offering a small set of intuitive and simply defined input handles by which an artist can author effects of their liking.  
To this end, throughout the method, we highlight the following four criteria that this method aims at preserving: (a) an explicit and comprehensible parameterization of the effects compatible with art-directability; (b) handles for manually tuning effects up to the vertex level; (c) preserving real-time computation in the order of magnitude of plain LBS; (d) ensuring stability and predictability of the resulting effect.

Consistent with extensive artist evaluation in our previous efforts~\cite{accelerationskinning,velocityskinning}, dynamic skinning fits into the animation pipeline in a practical way.  Namely, after skeletal animation is generated (and fixed), secondary motion can be added in a serial, controllable way to enhance the artist's intended effect on a particular animation.  In dynamic skinning, as effect parameters for oscillation and time-delay propagation are set for a given animation (e.g. scene or shot) they provide an excellent starting point for subsequent animation effects, while intuitive tuning knobs support the micro-adjustments each shot may need to fully realize an artist's vision.  This workflow is a more direct approach for adding secondary effects compared to the set and test efforts usually associated with physically based secondary tools because the technique is very fast and tuning parameters have a smooth, predictable response comparably.

Further, along with effects described by the velocity and acceleration skinning work, dynamic skinning offers a range of
deformations that can be combined
easily.  Also,
dynamic skinning supports ease of use through tools including the application of vertex weight painting that suppresses (or promotes) specific regions of the 
body, including unique animation 
within specific bodies of a skeleton.  Alternatively, we showcase a more simplified version of this control that is separated by body part and follows the LBS weights for smooth transitions between body parts. 
Results include examples of this in the accompanying video.
Beyond velocity and acceleration skinning, dynamic skinning provides hooks for generating more complex effects, like those of higher order derivatives, approaching the expected time response of an object in a physical setting.  As such, we see dynamic skinning as the capstone of the effort, giving an artist a fully featured tool for constructing physical secondary responses from ``primary'' input motion.  

In conclusion, in this paper, we present dynamic skinning as an extension of velocity skinning and acceleration skinning, which allows for time-dependent deformations based on joint kinematics. DS modifies the technique of Linear Blend Skinning (LBS) and extends the skeleton-based animation that is commonly used by artists for rigged-character animation and allows for ``on the fly'' adjustments of keyframe animation changes giving artists the ability to fine-tune and easily configure their animation while adding cartoon-like characteristics. DS extends upon velocity skinning and acceleration skinning methods allowing for the time dependent deformations based on a simple filtering scheme to allow for the new time-dependent effects, for which we have shown examples of such deformers: a primary oscillation effect, 
delayed oscillation, and delayed
LBS, as well as acceleration/deceleration asymmetrical behavior. 
Future work could involve defining custom implementation of time filters allowing for greater potential for other deformations allowing for custom kernels, and new ways to combine two or more deformers. The method is able to be implemented on GPU to handle bigger vertex count on massive meshes.

\section*{Acknowledgement}
We acknowledge Otman Benchekroun for his help with the comparisons with his work and fruitful discussion.
We acknowledge and thank Clemson University's Digital Production Arts program for giving us permission to use the character ``Doodle'' from the short film, \textit{Papel}. We also thank the production team behind \textit{Papel}: Nancey Fang, Insun Kwon, Harrison Diesl, Hang Yuan, Christine Wendell, Olivia Felber, Will Gilmer, Alexis Hunter, Ashley Johnson, Samuel New, and Julie Huang for envisioning and creating this character and short film.

\bibliographystyle{abbrv-doi-hyperref-narrow}
\bibliography{bibliography}

\begin{appendices}

\section{Computing the hierarchical weights}
\label{sec:hierarchical_weights}

Algorithm~\ref{algo:hierarchical_weights} describes a simple way to compute the hierarchical weights \(\beta_{ij}\) from the skinning weights \(\alpha_{ij}\).

\begin{algorithm}
\caption{Computing hierarchical weights $\beta_{ij}$}
\begin{algorithmic}
    \For{$i\in \I$}
    \State \mbox{Initialize} $\forall j\in \mathcal{J},\;\beta_{ij} = 0$
    \For {$j_0$ such that $\alpha_{i j_0}>0$}
        \State $\beta_{i j_0} = \alpha_{i j_0}$
        \State $j= j_0$
        \While{$j>0$}
            \State $\beta_{i a(j)} = \beta_{i j}+\alpha_{i a(j)}$
            \State $j= a(j)$
        \EndWhile
    \EndFor
    \EndFor
\end{algorithmic}
\label{algo:hierarchical_weights}
\end{algorithm}

\section{Integration by parts of Eq.~(\ref{eq:convolution2})}
\label{appendix:integration}

We show that 
\[
\displaystyle \int_{u=0}^{W} \dot{\kappa}(u)\,\Omega_j(t-u)\;\mathrm{d}u = 
\displaystyle \int_{u=0}^{W} \kappa(u)\;\,\dot{\Omega}_j(t-u)\;\mathrm{d}u
\]
Integrating by parts the left-hand side leads to
\[
\begin{array}{ll}
\displaystyle \int_{u=0}^{W} \dot{\kappa}(u)\,\Omega_j(t-u)\;\mathrm{d}u
=  & 
\underbrace{\bigl[\kappa(u)\,\Omega_j(t-u)\bigr]_{0}^{W}}_{\simeq 0} \\ & \displaystyle  - \int_{0}^{W} \kappa(u)(-\dot{\Omega}_j(t-u))\;\mathrm{d}u \mbox{ .}
\end{array}
\]
The first term is negligible as we assume \(\kappa(0)=0\), and \(\kappa(W)\simeq 0\), which leads to the desired result.

\section{Kernel alternatives}
\label{appendix:kernel}
We describe in the following three examples of possible custom procedural oscillation kernels \(\kappa_1, \kappa_2\), and \(\kappa_3\) illustrated in Fig.~\ref{fig:oscillation_type}.

First, we can represent a time-varying frequency by considering
\[
\kappa_1(t)=\sin\left(2\pi \left(f_0+f_s t\right) t\right), 
\]
where \(f_0\) represents the initial frequency and \(f_s\) is the slope of a linear frequency increase. This kernel models a vibration featuring low frequency at the start of the motion (i.e. large magnitude), and higher frequency at its end (i.e. low magnitude).  The effect of this mimics an object's response to disturbance which damps out but appears to do so with faster and smaller oscillations.

Second, we consider a kernel representing oscillation with long-lasting maxima mimicking the type of vibration observed on a pendulum with large amplitude. We formulate this effect by considering the first harmonics of a trapezoid as 
\[
\begin{array}{l}
\kappa_2(t)= \\ \frac{1}{1+1/9+1/25} \left(\sin(2\pi f t)+ \sin(6 \pi f t)/9-\sin(10 \pi f t)/25\right).
\end{array}
\]

Third, we consider the extreme case of a triangular response which conveys a sense of a more artificial, mechanical-like oscillation. In such a case, we set 
\[\kappa_3(t)=\mbox{triangle}(f,t), \] 
where \(\mbox{triangle}(f,t)\) is the triangle function of frequency \(f\) between \([-1,1]\).

\section{User-defined parameters}
\label{appendix:parameters}

We describe in this appendix the parameters and the default values that can be used to tune the results.

\subsection{Input data}

We first present in Table~\ref{table:input-data} the input data attached to the rigged model that represents the spatial extent of the deformation associated with the joints.

\begin{table}[h]

\caption{Input data associated with the model}

\begin{tabular}{cll}
\hline
\textbf{Parameter} & \textbf{Role} & \textbf{Value} \\
\hline
$\alpha_{ij}$ & Skinning weights & User defined (rig) \\
\hline
$\beta_{ij}$ & Hierarchical weights & Eq. (\ref{eq:hierarchical_weights}) \\
\hline
\end{tabular}
\label{table:input-data}
\end{table}

\subsection{DS default parameters}

Table \ref{table:parameters-default} presents the default parameters of the DS model. All these parameters can be modified straightforwardly by the user in order to adapt the general deformation and oscillation effects. We assume a model to be of unit size (1 m) for size-related parameter values.

\begin{table}[h]
\centering
\renewcommand{\arraystretch}{1.2}
\caption{Standard DS parameters, their roles, and typical values.}

\begin{tabular}{c p{3.5cm} c}
\hline 
\textbf{Parameter} & \textbf{Role} &  \textbf{Typical value} \\
\hline \hline
\multicolumn{3}{ l }{ \textbf{Bend effect $\psi^{\txt{bend}}$}, see Sec. \ref{eq:drag-effect}} \\
\hline
$\gamma^{\txt{bend}}$ & Bending magnitude & $1$ \\
\hline
\multicolumn{3}{ l }{\textbf{Oscillation kernel $\kappa$}, see Sec. \ref{sec:oscillation} } \\
\hline
$A$   & Oscillation magnitude    & $0.3$ \\
\hline
$\tau$ & Oscillation damping  & $1 \,(\text{s})$ \\
\hline
$f$   & Oscillation frequency    & $1.5 \,(\text{Hz})$ \\
\hline
$W$   & Convolution time window    & $5\, \tau (\text{s}) $ \\
\hline
\multicolumn{3}{ l }{\textbf{Time delay}, see Sec. \ref{sec:delay-filter}} \\
\hline
$c$   & Wave propagation speed    & $1$ (m/s) \\
\hline
\end{tabular}
\label{table:parameters-default}
\end{table}

\subsection{Advanced fine-tuning}

Finally, the DS deformation can be fine-tuned with the following possibilities:

\begin{itemize}
\item Adding a per-vertex and/or per-joint dependence to the parameters \(\gamma^{\txt{bend}}_{ij}, A_{ij}, \tau_{ij}, f_{ij}\), see Sec. \ref{sec:weight_painting}.
\begin{itemize}
\item A per-vertex dependence, that can be manually painted, allows fine-tuning the deformation beyond the level of precision of the rig.
\item A per-joint dependence allows for treating each joint-related deformation separately and relying on the smoothness and expressivity of the skinning weights.
\end{itemize}
\item Setting a varying deformation origin via the flexible origin operator \(O_{j_{\txt{min}}}\), allowing the representation of the effect of a flexible limb attached to a more rigid body part, see Sec. \ref{sec:deformation-origin-control}.
\end{itemize}

\newpage

\section{Pseudo code for DS computation}
\label{appendix:pseudo-code-DS}

We describe in Algorithm~\ref{algorithm:DS} the pseudo-code corresponding to the DS skinning deformation, including the use of the procedural oscillation effect and bend deformer.

\algrenewcommand\textproc{}%

\begin{algorithm}

\caption{Computing Dynamic Skinning deformation}
\begin{algorithmic}[1]

\State {\color{gray}\em \# Oscillation effect }
\Function{$\kappa$}{$t$} \hfill 
    \State \Return $A \exp\,(-t/\tau) \sin(2 \pi f t)$ \Comment{Eq.(\protect\ref{eq:kernel_sinus})}
    \State {\color{gray}\em \# Or any alternative kernel function as in Appendix \ref{appendix:kernel}. }
\EndFunction

\Statex

\State {\color{gray}\em \# Bend deformer}
\Function{$\psi^{\txt{bend}}$}{$\Omega,u$} \hfill 
    \State axis = \(\Omega/\|\Omega\|\), angle = $-\gamma^{\txt{bend}}\,\|\Omega \times u\|$
    \State build \(R=\)Rotation(axis,angle)
    \State \Return $(R-I)\;u$ \Comment{Eq.(\ref{eq:drag-deformer})}
\EndFunction

\Statex
\Statex

\State $\forall i\in \mathcal{I}, \forall j\in \mathcal{J},$ Initialize $\beta_{ij}$ \Comment{Appendix \ref{sec:hierarchical_weights}}
\State Initialize $\forall j\in \mathcal{J}$, $L_{\Omega_j} = [\,]$ \hfill {\color{gray}\em\# Stored angular velocity}

\State {\color{gray}\em\# Main animation loop}
\For{$m=0$; while true} 
    
    \State Initialize $\forall j\in J,\; L_{\Omega_j [m]} = \Omega_j$
    \State $\forall i\in I,\;p_i^{\LBS} = \text{LBS}(p_i)$ \hfill {\color{gray}\em\# Linear Blend Skinning}

    \ForAll{$i \in I$}
        \State Initialize $\mbox{DS}_{\txt{{deform}}} = 0$
        \ForAll{$j \in J$}

            \State {\color{gray}\em\# Numerical convolution}
            \State $\Omega_{\txt{filtered}} = L_{\Omega_j [m]}$
            \For{$n=0;$ while $ n < W / \Delta t$}
                \State $\Delta \Omega_j = L_{\Omega_j [m-n]} - L_{\Omega_j [m-n-1]}$
                \State $\Omega_{\txt{filtered}} \;+\!\!=  
                \kappa(n \Delta t) \; \Delta \Omega_j$ \Comment{Eq.(\ref{eq:convolution_numerical})}
                \State $n=n+1$
            \EndFor
            \State {\color{gray}\em\# DS deformation for joint $j$ at vertex $i$}
            \State $\mbox{DS}_{\txt{{deform}}} \;+\!\!= \beta_{ij} \; \psi^{\txt{bend}}(\Omega_{\txt{filtered}}, p_i^{\LBS} - q_j)$ 
        \EndFor
        \State $p_i^{\LBS} = p_i^{\LBS} + \mbox{DS}_{\txt{{deform}}}$ \Comment{Eq.(\ref{eq:dynamic-skinning})}
    \EndFor

    \State $m = m+1$
\EndFor

\end{algorithmic}
\label{algorithm:DS}
\end{algorithm}

\end{appendices}

\end{document}